\documentclass[manuscript]{aastex}

\shorttitle{GRB980923. A burst with a short duration high energy component.}
\shortauthors{M. M. Gonzalez, J. R. Sacahui, J.L. Ramirez, B. Patricelli and Y. Kaneko}

\begin{document}


\title{GRB980923. A burst with a short duration high energy component.}


\author{M. M. Gonz\'alez}
\affil{Instituto de Astronom\'\i a, UNAM,
    M\'exico, 04510}

\author{J. R. Sacahui}
\affil{Instituto de Astronom\'\i a, UNAM,
    M\'exico, 04510}

\author{J.L. Ramirez}
\affil{Instituto de Astronom\'\i a, UNAM,
    M\'exico, 04510}
    
   \author{B. Patricelli}
\affil{Instituto de Astronom\'\i a, UNAM,
    M\'exico, 04510, \\ICRANet, Pescara, Italy, I-65100} 
    
   \author{Y. Kaneko}
\affil{Sabanci University, Orhanli-Tuza,
    Istambul, Turkey, 34956 } 




\begin{abstract}
The prompt emission of Gamma Ray Bursts (GRBs) is usually well described by the Band function: two power-laws joined smoothly at a given break energy. In addition to the Band component, a few bursts (GRB941017, GRB090510, GRB090902B and GRB090926A) show clear evidence for a distinct high-energy spectral component, which in some cases evolves independently from the prompt keV component and is well described by a power-law (PL), sometimes with a cut-off energy; this component is found to have long duration, even longer than the burst itself for all the four bursts. Here we report the observation of an anomalous short duration high energy component in GRB980923. GRB980923 is one of the brightest Gamma-Ray Bursts (GRBs) observed by BATSE. Its light curve is characterized by a rapid variability phase lasting $\sim$ 40 s, followed by a smooth emission tail lasting $\sim$ 400 s. A detailed joint analysis of BATSE (LAD and SD) and EGRET TASC data of GRB980923 reveles the presence of an anomalous keV to MeV component in the spectrum that evolves independently from the prompt keV one. This component is well described by a PL with a spectral index of $-1.44$ and lasts only $\sim$ 2 s; it represents one of the three clearly separated spectral components identified in GRB980923, the other two being the keV prompt emission, well described by the Band function and the tail, well fit by a Smoothly Broken Power Law (SBPL). 
\end{abstract}





\section{Introduction}\label{sec:intro}

Gamma-Ray Bursts (GRBs) are the most energetic sources in the universe, with a total isotropic equivalent radiated energy in the range $10^{49} - 10^{55}$ erg. They are characterized by a brief and intense flash, the so-called prompt emission, observed in  $\gamma$ and X-ray energy bands, followed by a long lived afterglow emission, radiated in the X-ray band and below. The GRB prompt emission spectra are typically best fit by the ``Band function": two power-laws joined smoothly at a given break energy \citep{B93}, whose low energy and high energy photon indices, $\alpha$ and $\beta$, have median values of -1 and -2.3 respectively \citep{PR00,K06}. In addition to the Band component, a few bursts show clear evidence for a distinct high-energy (HE) spectral component. The first GRB showing such a characteristic was GRB941017, for which an extra multi-MeV spectral component has been observed \citep{GO03}. This component lasts longer (200 s) than the  keV component (T90\footnote{The T90 is defined as the time during which the cumulative counts increase from 5\% to 95\% above background, thus encompassing 90\% of the total GRB counts \citep{Kouv93}.} $=$ 77 s), from which it evolves  independently in time; it is well fit by a PL with spectral index -1.0 and carries two thirds of the energy fluence. 

With the launch of the {\it Fermi} gamma ray space telescope \citep{Atwood09,Megaan09} other three bursts with a clear distinct HE spectral component besides the Band function have been observed: GRBs 090510 \citep{Ackermann090510}, 090902B  \citep{Abdo090902B} and  090926A \citep{Ackermann090926A}. This component  is usually well fit by a hard power-law that dominates at higher energies \citep{Granot10} and has a long duration, longer than the duration of the burst itself. 

Specifically, GRB090510, characterized by $T_{90}=2.1$ s, presents an extra PL component with a spectral index $-1.62$ during the prompt phase; this short duration HE emission was delayed by $\sim$0.1 s with respect to the onset of the Band component and was followed by an extended HE emission,  lasting $\sim$ 200 s and well fit by a PL \citep{Ackermann090510,dePasquale2010}. 

GRB090902B, characterized by $T_{90}=21.9$ s, shows a hard additional component during the prompt phase which is well fit by a single PL with spectral index $-2.1$ and lasts $\sim$ 1000 s. Furthermore, a spectral feature at energies $<$ 50 keV is evident, that is consistent with an extrapolation of the HE PL emission down to those energies \citep{Abdo090902B}.

The last {\it Fermi} burst, GRB090926A, shows an extra HE component that is very significant at the time of a narrow pulse (lasting less than 1 s) observed in the prompt emission light curve; the HE emission is also present at later times, with an overal duration of hundreds of s (see also sec. \ref{sec:band}). The fit of the narrow pulse spectrum requires, as seen for the other bursts, an extra hard PL component besides the Band one, but for the first time it has been detected a spectral break in this PL around 1.41 GeV. After this pulse the HE emission is well described by a simple PL \citep{Ackermann090510}.

There are two main classes of models that have been proposed to describe the HE  $\gamma$-ray emission: leptonic and hadronic. Between the leptonic models, the most investigated scenarios consider Inverse Compton (IC) and Synchrotron Self-Compton (SSC) emission processes in different locations of the relativistic jet from which the GRB is generated. Two regimes for accelerating electrons are considered: they can been accelerated in external (as the jet interacts with the circumburst medium) and internal (within the jet as the Lorentz factor of the flow varies) shocks; different seed photon populations for the IC scattering has been discussed in detail in GRB internal shocks \citep{pap96,pil98, pan00a},  forward shocks \citep{sar96,tot98,wax97,pan98,wei98,chi99,der00a,der00b,pan00b} and  reverse shocks \citep{wan01a,wan01b, pee06}.  

In particular,  SSC processes  from forward  \citep{sar01,wan01a} and reverse shocks \citep{Granot03,wan01a,wan01b} have been investigated  separately to explain some anomalous HE components such as the one of GRB941017. Also, Synchrotron emission propagating into an ambient with a sufficiently low density has been proposed to explain the HE emission observed in GRB090510 \citep{he11} and GRB090902B \citep{liu11}.  More recently,  \citet{fra12, ver12} explored SSC processes from forward and reverse shocks as possible mechanisms to produce the HE emission and concluded that, depending on the equipartition parameters for the magnetic field and the electron energy in the forward and reverse shock, one or the other could develop to explain the HE component. 

Within the hadronic models, the two most investigated scenarios consider respectively: 1) Synchrotron radiation from ultra HE protons accelerated in the relativistic jet \citep{vietri97,tot98,tot98b,razzaque09,razzaque09b}; 2) Synchrotron and IC emission from secondary electron-positron pair cascades triggered by photopion interactions of shock accelerated protons with low energy photons coming from the prompt emission or from an external radiation field \citep{wax97b,bottcher98}. \citet{asano09} investigated both the scenarios to explain the HE emission observed in GRB090510 finding that, in both cases, to produce the HE component of this burst the proton injection isotropic-equivalent luminosity is required to be larger than $10^{55}$ erg/s, thus larger than the $\gamma$-ray luminosity; in the case of GRB090902B, however, the keV-GeV spectrum detected in GRB 090902B is well explained with a comparable energy in protons and $\gamma$-rays \citep{asano11}. 

An additional process has been proposed to explain the HE emission observed in GRB941017 by \citet{dermer04}. They considered a beam of ultra HE neutrons produced in the photopion interactions of ultra-relativistic protons with internal and external radiation photons; this neutron beam can undergo further interactions with external photons to produce hyper-relativistic electrons that radiate energy by Synchrotron, producing the anomalous HE component. 

Most of the models described consider an extra HE component that has a long duration in comparison with the burst $T_{90}$ and that is well described by a hard PL, with a photon index almost constant in time.   

Within this context an interesting case of study is GRB980923, one of the brightest bursts detected by the Burst And Transient Source Experiment (BATSE), characterized by an anomalous short duration HE emission during the prompt phase. Some evidence for the presence of this extra HE component was first reported by  \cite{GO05} who analysed joint data from the BATSE Large Area Detector (LAD) and the Energetic Gamma Ray Experiment Telescope (EGRET) calorimeter, the Total Absorption Shower Counter (TASC), finding that an extra PL, besides the Band function, is needed to correctly reproduce the spectrum of this burst from 13 s to 46 s. Here we extend the spectral analysis of GRB980923 using, besides the data from the BATSE-LAD and EGRET-TASC, data from the BATSE Spectroscopy Detector (SD), gaining time-resolution and energy coverage (see sec. \ref{sec:analysis}): this allow us to determine with more accuracy the time duration, as well as the spectral index of the HE  component. In particular, time-resolved spectral analysis revealed that this HE component only lasts  $\sim$ 2 s, in contrast with the overall long duration of the extra HE component of the other GRBs. This component represents one of the three clearly separated spectral components identified in GRB980923, the other two being the prompt emission, well described by the Band function and a tail, lasting 400 s and well fit by a PL \citep{GI99}.  

The paper is organized as follows. In sec. \ref{sec:grb980923} we summarize the main observational properties of GRB980923. In sec. \ref{sec:analysis} we briefly describe BATSE and EGRET-TASC and the set of data chosen; we also explain how the analysis of these data was performed. In sec. \ref{sec:results} we present the results. In sec. \ref{sec:conclusions} we present our conclusions.

\section{GRB980923}\label{sec:grb980923}

GRB980923 was observed by BATSE at 20:10:52 for 33.02 s (as determined by BATSE T90) from 
the galactic coordinates $l=293.08^\circ$ and $b=-30.89^\circ$. The burst direction was 
$124.4^\circ$ with respect to the pointing-axis direction of the Compton Gamma Ray Observatory (CGRO), placing the burst outside the $\sim$1 sr field of view of the Imaging Compton Telescope (COMPTEL) and the EGRET spark chamber. With a fluence of $4.5\times 10^{-4}$ erg/cm$^{2}$, GRB980923 is the burst with the third highest fluence  in the BATSE catalog \citep{PR00}.

The prompt emission lightcurve of GRB980923  as observed by BATSE-LAD and BATSE-SD has two main episodes (see Figure \ref{LCs_980923}, upper panels). The first episode lasts about $\sim$14 s, then the count rate drops drastically, for about 2 s, to start the second episode, whose duration is of $\sim$20 s. A single Band function is not sufficient to describe the whole prompt emission from 20 keV to 200 MeV, as shown in the joint analysis of data from BATSE-LAD and EGRET-TASC performed by  \citet{GO05, KA08} and \citet{GO09}.  In particular, \citet{GO05} found that to fit the second episode and extra PL is needed.

The prompt emission of GRB980923 is followed by a long and smooth tail, whose duration is reported to be $\sim$400 s \citep{GI99}. This tail is specifically studied by \citet{GI99} using BATSE data. Its time-integrated spectrum is well described by a Smoothly Broken Power Law (SBPL), as well as by the Band function. \citet{GI99} identified a separate emission component related to the tail starting at $\sim$ 40 s when the spectral shape changes dramatically. The temporal evolution of the energy break of the SBPL showed that if the tail starts at 32.109 s after the trigger, the evolution of its spectrum corresponds to the evolution of a synchrotron cooling break in the slow-cooling regime and the transition from fast cooling could happen on timescales comparable to the duration of the burst. Therefore, if the synchrotron emission from external shocks explains the tail then, at least for some bursts, the afterglow may begin during the prompt phase \citep{GI99}. 

\section{BATSE and EGRET-TASC data}\label{sec:analysis}
In this work, data from BATSE and EGRET-TASC were used. A brief description of these instruments and of the set of data chosen is given below, together with some details about how the  spectral analysis of the data was performed.

BATSE (for a detailed description see  \citet{FI89, PE95,PR00} and \citet{K06}) consisted of eight separated identical detector modules located at the corners of the CGRO spacecraft on the faces of a regular octahedron. Each module had two NaI(Tl) scintillation detectors coupled with photomultiplier tubes (PMTs): a Large Area Detector (LAD), optimized for sensitivity and directional response and a Spectroscopy Detector (SD), optimized for energy coverage and resolution. Each detector had different energy capabilities; LAD had a constant energy range of 0.02 - 1.9 MeV while SD had an adjustable energy range between  0.01 and 100 MeV depending on the PMTÕs gains. Even though the SD was sensitive over a broader energy range, the LAD had a collecting area 16 times bigger than the SDÕs. Each detector module was independent. Data were processed and accumulated to construct various data types with different energy resolutions and accumulation times in the data processing unit. In the analysis of \citet{GO03, GO05} only LAD data were used because of the bigger collecting area. In this paper, we also include SD data to increase the energy range with finer time resolution than EGRET-TASC. In the case of LAD, we use data from detector 7 (LAD7) in the Continuous (CONT) format that contains 16 energy channels with an accumulation time of 2.048 s. In the case of SD, we used data from detectors 3 (SD3) and 7 (SD7) in the Spectroscopy High Energy Resolution Burst (SHERB) format, that contains spectra in 256 energy channels with an accumulation time of 512 ms. SD7 and SD3 data allowed us to extend the analysis to lower energies (32 keV)  and to higher energies (27 MeV) respectively.

TASC is a monolithic 76 x 76 x 20 cm$^3$ scintillation calorimeter formed by 36 NaI(Tl) blocks optically coupled. It was viewed by two groups of eight interleaved PMTs, each feeding a pair of Pulse Height Analyzers (PHAs). One pair of PHAs processed low-energy events (1 - 200 MeV), while the other processed high-energy events (0.02 - 30 GeV). Although the TASC was part of the EGRET instrument measuring the energy of each useful event triggering the spark chambers
(mainly done with the high-energy PHA), it was also an independent detector sensitive to gamma rays and charged particles (using the low-energy PHA). As an independent detector, the accumulation time of 32.768 s for low-energy spectrum was the normal and continuous mode of data acquisition and was called the solar mode. The TASC data used for GRB980923 consists of only two solar spectra with 229 energy channels, starting 19.6 s before the BATSE trigger and covering almost twice the burst duration (for a detailed description see \citet{GO09}). 

\begin{figure}
      \epsscale{0.7}
      \plotone{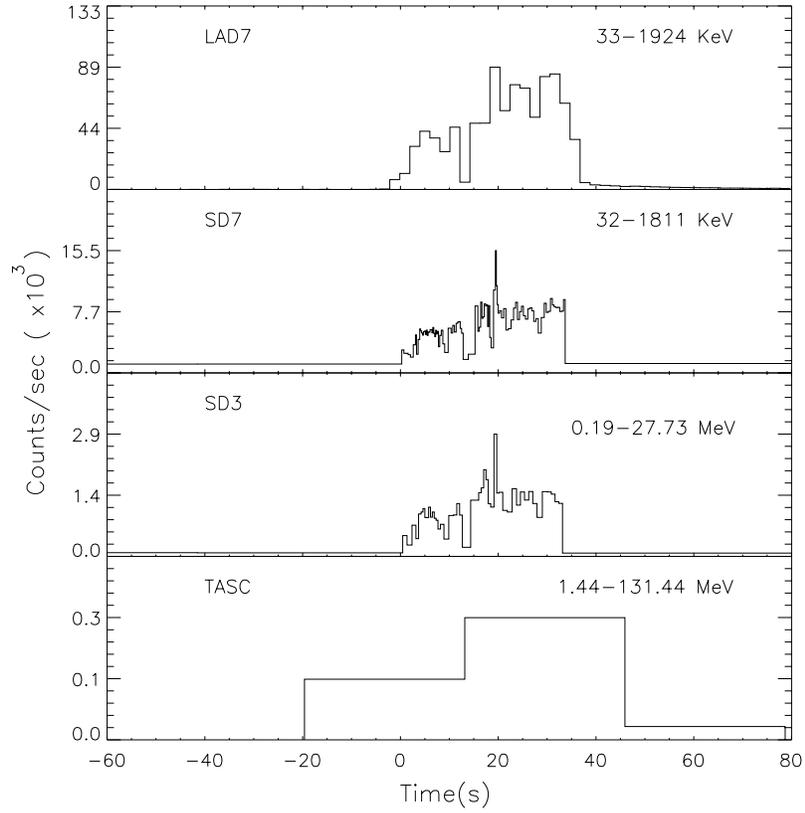}
      \caption[Total Lightcurves for GRB980923]
{Count rates for GRB980923. Lightcurves for LAD, SDs and
  TASC detectors are shown. BATSE count rates decrease considerably around 14 s. SD lightcurves present a sharp peak around 20 s. The cutoff at 33 s is due to the ending of the data accumulation and not intrinsic to the burst lightcurve. A long tail is also evident in LAD data.}
\label{LCs_980923}
\end{figure}

The lightcurves of the emission observed by LAD7, SD7, SD3 and TASC are shown in Figure \ref{LCs_980923}. General features are seen in all the BATSE lightcurves: two main episodes separated by a drop in the count rate at $\sim$14 s with a duration of $\sim$2 s (see also sec. \ref{sec:grb980923}). The drop is not observed in TASC data because of the 32.768 s accumulation time of the spectra. The finer time-resolution of the SD7 and SD3 data allows to observe the structure of both episodes. In particular an intense peak at $\sim$20 s is more evident from the lowest to the highest BATSE energies. The long tail is evident in LAD data. 

The spectral fitting was performed with the software developed by the BATSE team, RMFIT\footnote{R. S. Mallozzi, R. D. Preece \& M. S. Briggs, ÒRMFIT, A Lightcurve and Spectral Analysis Tool,Ó \copyright Robert D. Preece, University of Alabama in Huntsville}. For a detailed description of the spectral analysis see \citet{GO09}; here, only relevant details to the analysis of GRB980923 are mentioned. An important step of the analysis is the background subtraction. The time dependency of the background was given mainly by the spacecraft position with respect to the Earth's magnetic field. We modeled background counts with a polynomial of 4th order on each energy bin using spectra for $\sim$ 500 s before and after the burst trigger. The background model was checked against data corresponding to 15 orbits
earlier/later when the spacecraft was located at the same geomagnetic rigidity. 
Background subtracted spectra of LAD7, SD3, SD7 and TASC data were fitted using the detector's responses. When a joint fit of BATSE and TASC data were performed, BATSE data were binned in time to match the time resolution of TASC data; when only BATSE data were used, SD data were binned to match the time resolution of LAD data. 
The energy ranges used in the fit for each data set are the same given for the lightcurves, see Figure  \ref{LCs_980923}.
The spectral fitting was performed with the following photon flux models: Band function, Band function plus a PL and a SBPL. 
As in \citet{GO09}, normalization factors between data sets were introduced to account for errors in the calculated effective area of the detectors because underestimation of the CGRO mass model. The value of the normalization factor depends mostly on the zenith angle of the burst. If the detector responses were perfectly known, the normalization factor would be one. The factors normalize TASC and SD data with respect to LAD data and it is the same for all energies in a same data set. However, in our analysis it varies from time bin to time bin because it also accounts for the mismatching of time binning between data sets. These normalization factors had similar values for both SD detectors. The TASC normalization factor is higher as expected mainly because of the greater accumulation time of TASC with respect to the other  detectors (see sec. \ref{sec:results}).

\section{Results and Discussion}\label{sec:results}
We performed the fit of prompt emission spectra of GRB980923 integrated over different intervals of time: this allowed us to identify and characterize with great accuracy the extra HE component of this burst, as well as to find some evidence of an early starting time of the tail.

\subsection{Spectral fit and identification of the HE component}\label{sec:band}
\begin{figure}
    \epsscale{0.75}
      \plotone{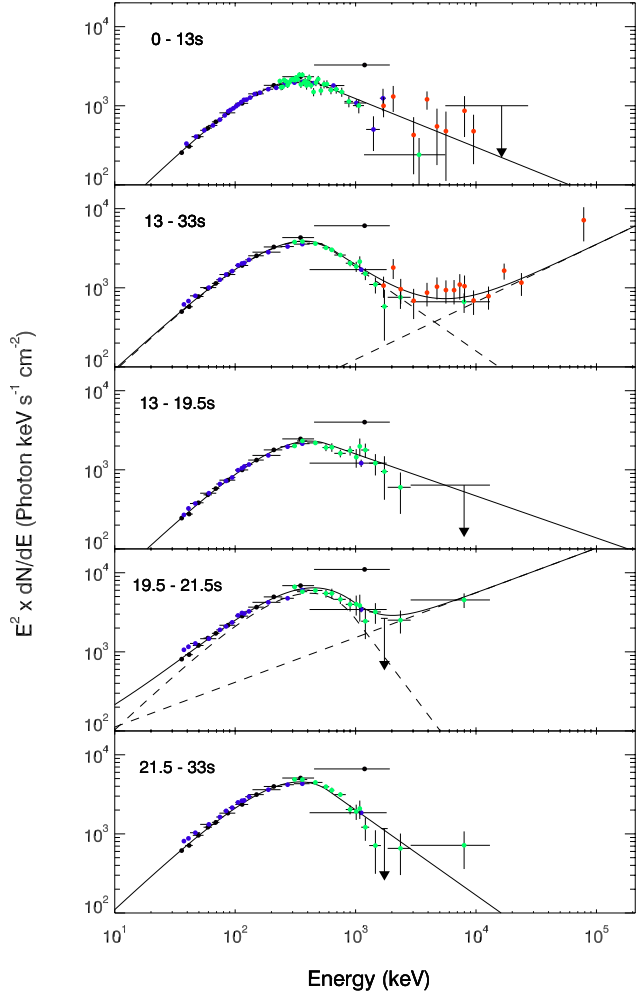}
      \caption[Spectrum of GRB980923]{Spectral fitting for different time intervals as defined in sec. \ref{sec:results}. LAD7 (black), SD7 (blue), SD3 (green) and TASC (red) are shown. For the purpose of the plot, but not for the spectral fit, the data are binned in energy to give at least 2$\sigma$ significance over background. Solid curves show model fits to the data using the parameters given in Table \ref{tbl980923} and the spectral model described in the text. The upper limits correspond to 2$\sigma$ deviation from the background. The two spectral components, the Band function at lower energies and the higher-energy power law are shown as dashed lines.  The intervals excluding the time interval from 19.5-21.5 s are well adjusted only with a Band function, otherwise a PL function is required to fit the highest energies.}
      \label{Ajustes}
      \end{figure}

We fit jointly LAD7-SD3-SD7-TASC data for the time intervals 0-13 s and 13-33 s determined by BATSE. The first time interval is fitted with a Band function, while the second one requires also a HE-PL function described by $\rm A_{PL}[E(keV)/10MeV]^\gamma$. The data and the spectral fits are shown in Figure \ref{Ajustes} (upper panels), with the best-fit parameters and the normalization factors given in the first two columns of Table \ref{tbl980923}. The normalization factors between LAD and SD detectors for both time intervals are very similar because the time coverage almost matches. 
However, the TASC normalization factors are different mainly because they account for the 18 s and 12 s of extra TASC time coverage compared to
LAD time coverage in the first and second time interval respectively. Also, the normalization factors are calculated considering the best fit to all data sets simultaneously. This and the contribution of the power law (when required)  are small when compared with the corrections for the time bin mismatching but they explain the difference with the values given by \citet{GO09}. The energy dependency, as well as the correct election of the normalization factors is confirmed by the fact that SD and TASC data below 10 MeV
smoothly continue the lower-energy component described by the Band
function in both time intervals (see Figure \ref{Ajustes}).  
 
\begin{figure}
      \epsscale{0.7}
      \plotone{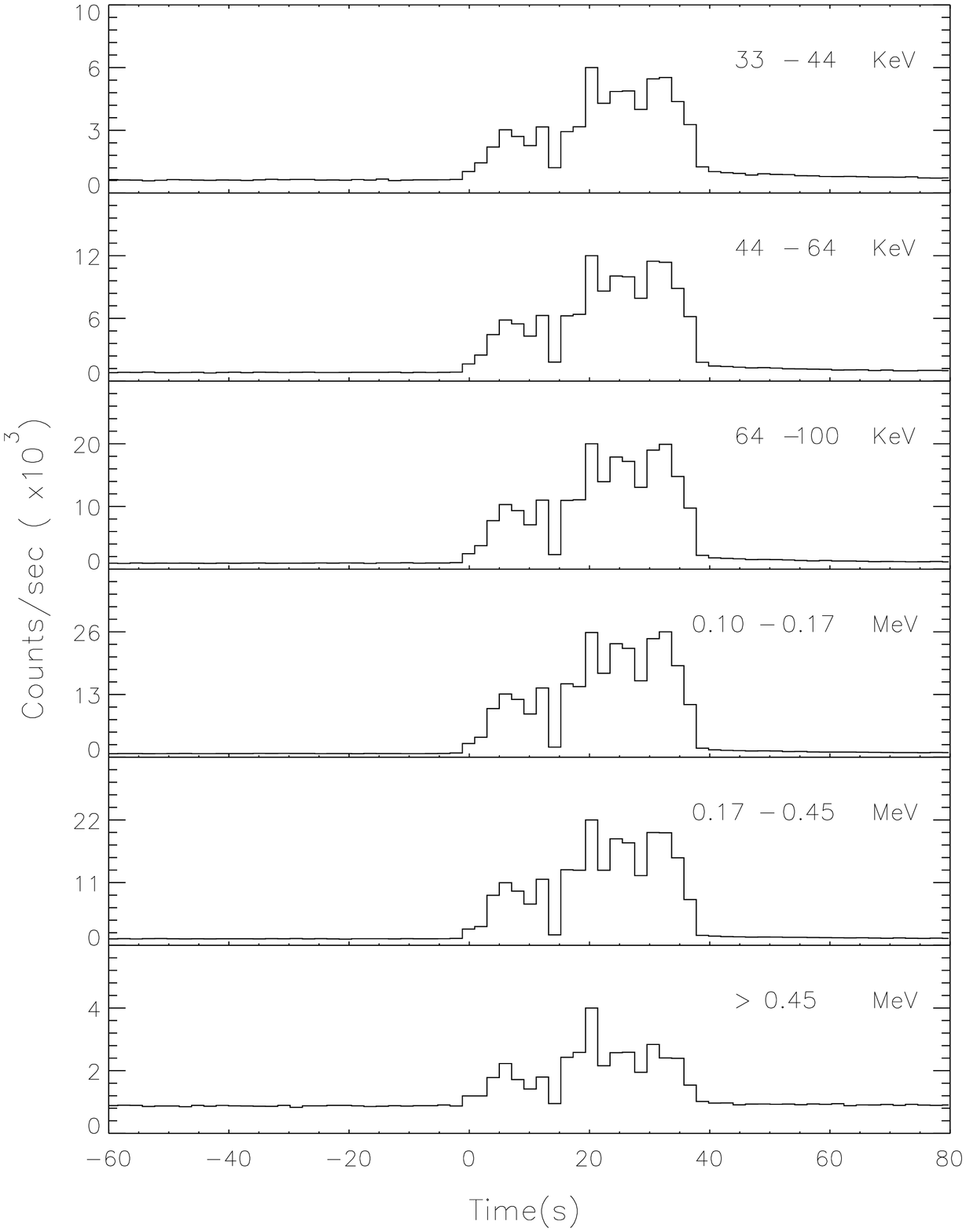}
      \caption[LAD7 Lightcurve for GRB980923 as function of energy]
      {LAD7 light curves in different energy bands (energy increasing from top to bottom). A peak at 20 s becomes more evident at the highest energies.}
\label{LAD7}
\end{figure}

\begin{figure}
      \epsscale{0.7}
      \plotone{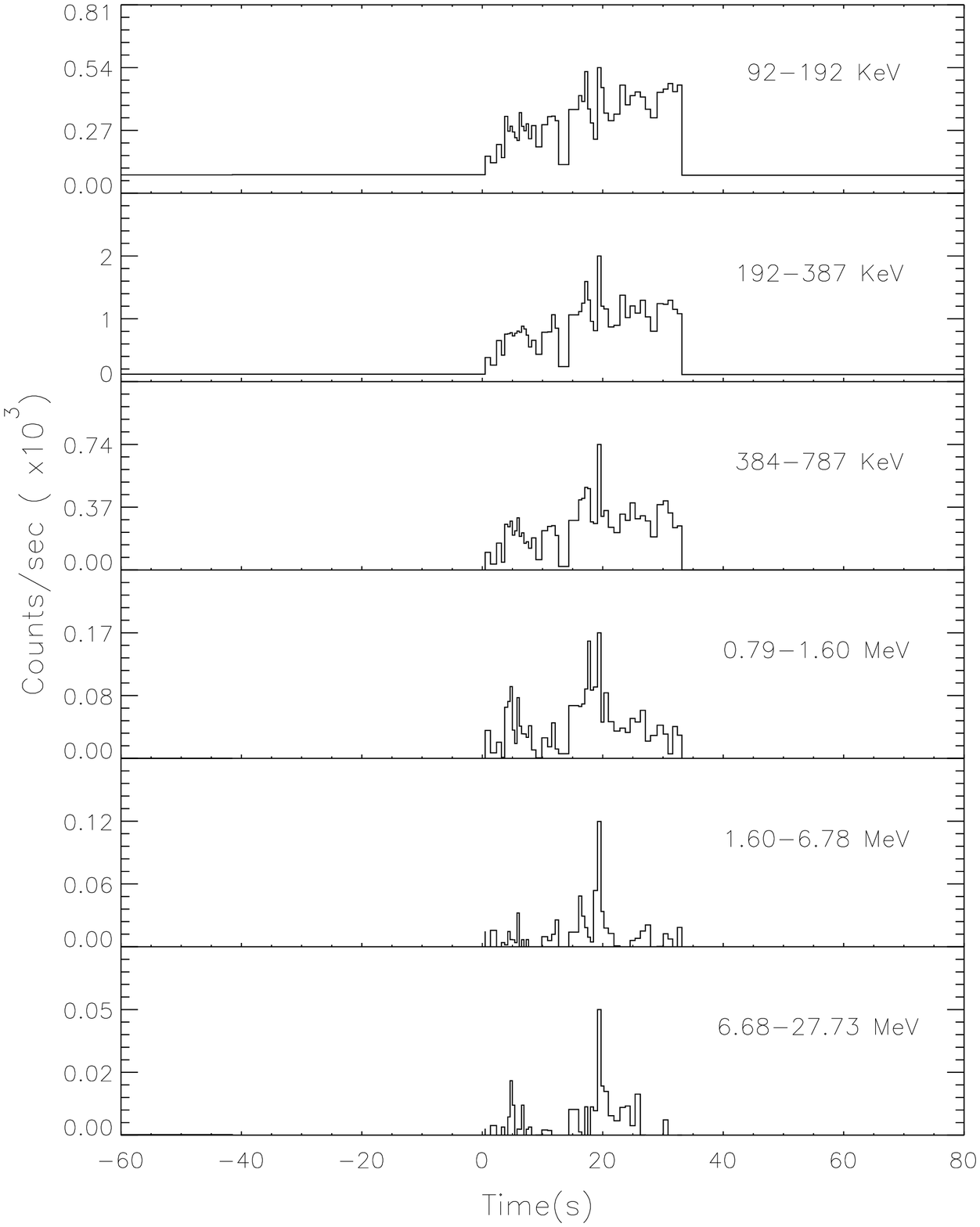}
      \caption[SD3 Lightcurve for GRB980923 as function of energy]
{SD3 light curves in different energy bands (energy increasing from top to bottom). A peak at 20 s becomes more evident at the highest energies.}
\label{SD3}
\end{figure}

\begin{figure}
      \epsscale{0.7}
      \plotone{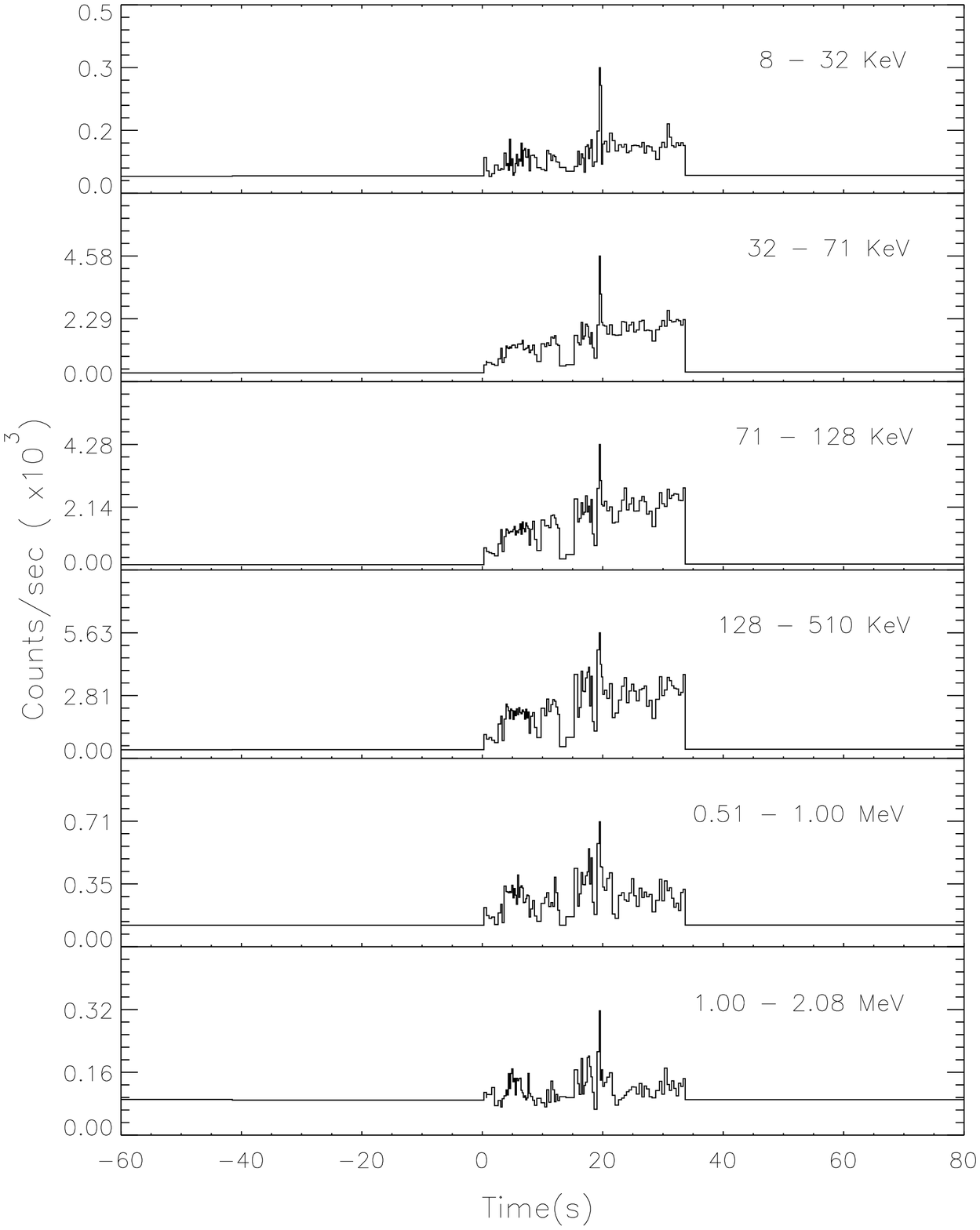}
      \caption[SD7 Lightcurve for GRB980923 as function of energy]
{SD7 light curves in different energy bands (energy increasing from top to bottom). A peak at 20 s becomes more evident at the lowest and highest energies.}
\label{SD7}
\end{figure}

\begin{figure}
      \epsscale{0.7}
      \plotone{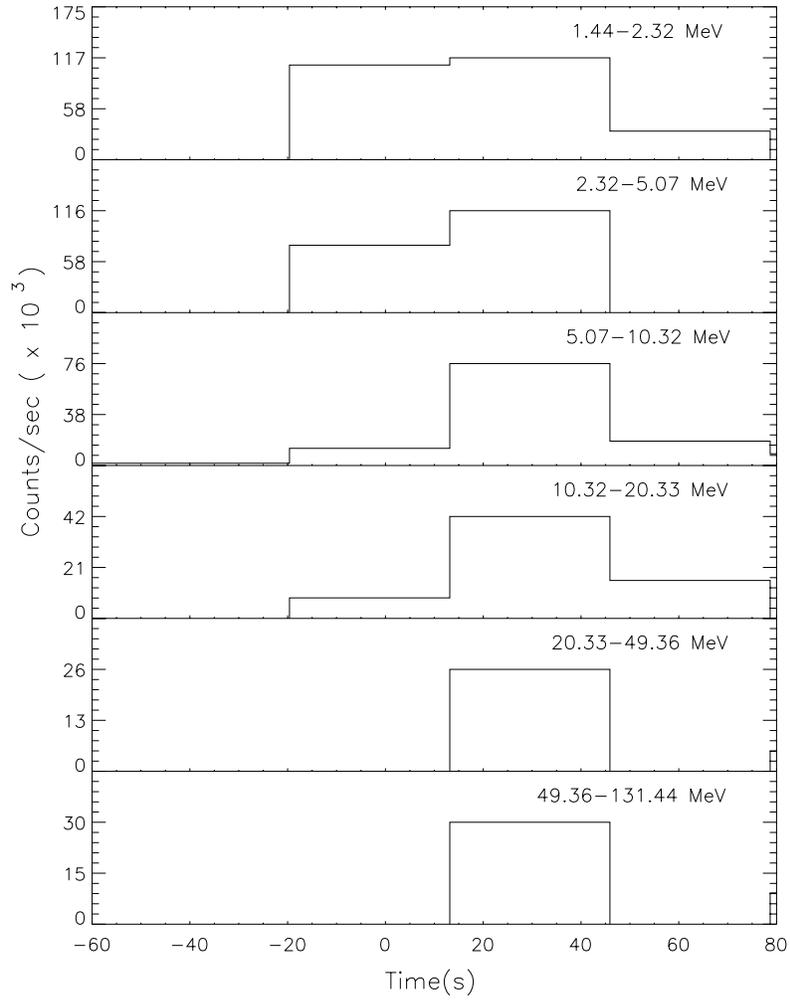}
      \caption[TASC Lightcurve for GRB980923 as function of energy]
{TASC light curves in different energy bands (energy increasing from top to bottom). The time interval including the peak at 20 s, as observed by BATSE, remains significant up to hundred of MeVs.}
\label{TASC}
\end{figure}

In order to better characterize the HE component and to determine with more accuracy its duration, we look at the lightcurves as function of energy for all detectors,
see Figures \ref{LAD7}, \ref{SD3}, \ref{SD7} and \ref{TASC}. The peak
at $\sim$20 s is evident from 8 keV up to 28 MeV as observed by 
SD7 and SD3 respectively. It evolves with energy differently from 
the whole lightcurve, that dims at the higher energies. This is consistent
with the TASC data: the first time interval dims at the higher energies, contrarily to the second one,  that includes the HE component. We
fit jointly LAD-SD data for 2 s time intervals from 0 to 33 s and find 
than only the time interval corresponding to the peak from 19.5 to
21.5 s requires a PL besides a Band function to fully
describe the data. The data and the spectral fits are shown in Figure \ref{Ajustes} (lower panels), with the best-fit parameters given in Table \ref{tbl980923}; the time evolution of the Band spectral indices is shown in Figure \ref{Index}. It can be seen that, except for the value of the low energy spectral index $\alpha$ at $\sim 14$ s and within the error bars, a smooth time evolution is observed in both the Band spectral
indices through the whole time period from 0 to 33 s. This suggests that the prompt emission described by the Band function comes from the same radiation process; the different value of 
$\alpha$ at $\sim 14$ s could indicate the presence of a distinct component starting during the prompt phase (see sec. \ref{sec:tail}) which could be noticeable because of the dim of prompt photons at $\sim 14$ s (see the three upper panels in Figure \ref{LCs_980923}). 
Concerning the PL component, the value of
the probability that the improvement in $\chi^2$ from the addition of the high-energy power law in the fit is due to chance, as determined by the $ \chi^2$ test considering the time interval from 19.5 to 21.5 s is four orders of magnitude (6 when TASC data are included) smaller in comparison with the one obtained for the time interval between 13 s and 33 s (see Table \ref{tbl980923}). We also fit smaller time
intervals included in the period from 19.5 s to 21.5 s using only SD
data, finding that the smallest probability that the improvement is due to change is found when considering the whole
2 s time period: this is an indication of the fact that the PL was always required along the 2 s time period. Therefore, the HE  component of GRB980923, extending from a few keVs to hundreds of MeVs, starts at 19.5 s and lasts only 2 s: GRB980923 represents then the first case of a burst showing an extra HE PL component of short duration (see Table \ref{tab:tail} for a comparison with the other bursts). The energy flux of this HE component is 
22 $\rm \times 10^{-6} erg\ s^{-1} cm^{-2}$ in the energy range from 2 to 200 MeV.

A burst that presents some similarities with GRB980923 is GRB090926A. Also this source is characterized by a HE emission associated with a short spike (lasting less than 1 s) observed in the prompt emission light curve, although a HE emission is also observed after the spike (see sec. \ref{sec:intro}). The HE emission associated to the narrow pulse is described by a PL having a different spectral index with respect to the one observed at later times and extends to the lowest enegies similarly to GRB980923. Moreover, a spectral break and the extension to the lowest energies
 are evident only when fitting the spike spectrum.
These observational properties suggest that we are in presence of two distinct HE components: one of short duration, that could have been produced by the same physical mechanism responsible for the HE emission observed in GRB980923 and one of long duration  \citep{sacahui12}. Other common characteristics between GRB980923 and GRB090926A are a drop in the count rate observed before the spike and the presence of a tail (see Table \ref{tab:tail}). 

\subsection{Spectral fit and implications for the tail}\label{sec:tail}
\begin{figure}
      \epsscale{0.7}
      \plotone{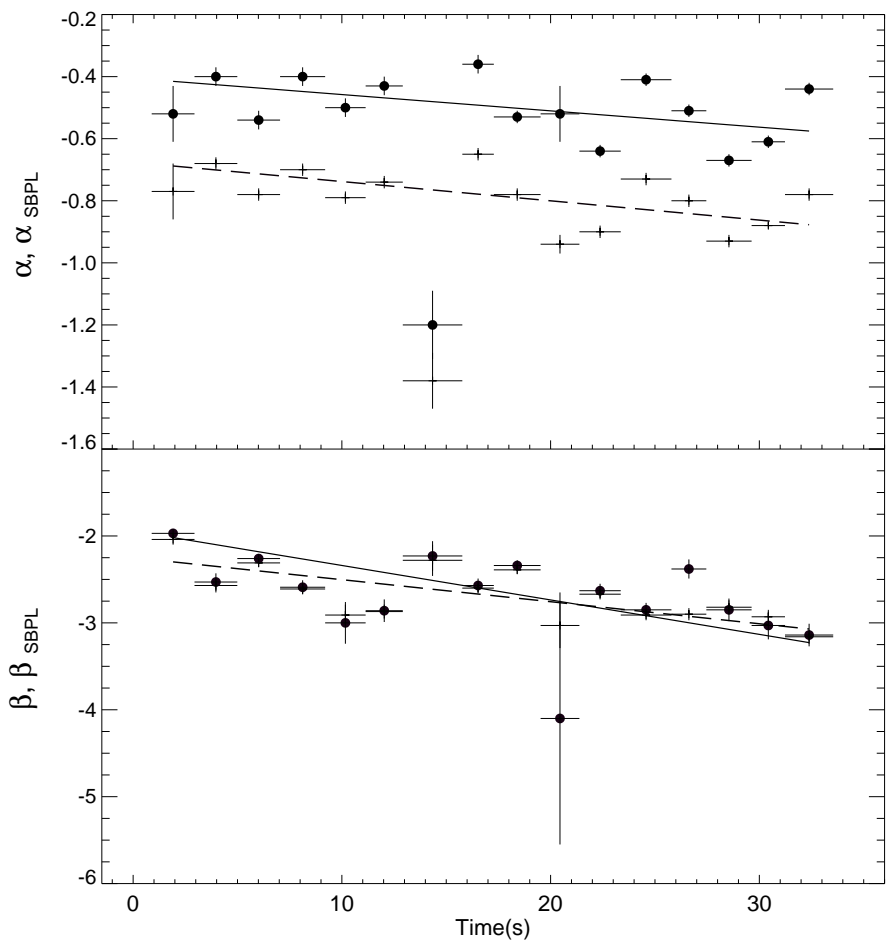}
      \caption[Spectral indices for the main prompt episode.]
{Spectral Indices as a function of time for the main prompt episode. Time evolution of the low and high energy indices for Band ($\alpha$ and $\beta$, solid circles) and SBPL ($\alpha_{SBPL}$ and $\beta_{SBPL}$, dots) functions. The solid and the dashed lines represent the best fit with a PL for the Band and the SBPL indices respectively. A smooth continuous evolution in time is observed from 0 s to 33 s, except at 14 s when the values of $\alpha_{SBPL}$ softens, becoming consistent with the values observed in the tail \citep{GI99}.}
\label{Index}
\end{figure}

We do the fit of time-resolved prompt emission spectra also using a SBPL for comparison with \citet{GI99}. Because of the low statistics at energies above 1 MeV, SBPL seems to be preferred by \citet{GI99}. However, we do not find any preference in the data to use SBPL or Band. The following discussion is independent of the fitting function used. The time evolution of the low energy and high energy spectral indices for the BAND and SBPL functions ($\alpha$, $\alpha_{SBPL}$, $\beta$ and $\beta_{SBPL}$ respectively) are shown in Figure \ref{Index}. It can be seen that the values of $\alpha$ and $\alpha_{SBPL}$  at 14 s do not follow the general trend. In particular, the value of $\alpha_{SBPL}$ is consistent with the one of the tail as given by \citet{GI99}, a possible indication of the presence of the tail already at 14 s. Furthermore, it is noticeable a higher dispersion of
the $\alpha$ and  $\alpha_{SBPL}$ values after $\sim$14 s (see Figure \ref{Index}) and this could be interpreted as a further evidence of the presence of the tail from 14 s. In fact, the time-integrated spectrum of the tail can be described with a SBPL with break energy around 200 keV \citep{GI99}; therefore, if the tail is present from 14 s, it will affect the determination of the low energy spectral index of the Band function, that has a peak energy around 400 keV. So it is possible that the tail starts at $\sim$ 14 s from the burst trigger instead of 32.109 s as reported by \citet{GI99} (see also sec. \ref{sec:grb980923}).  In this case, its behaviour is still similar to that of the afterglows at lower energies, but the new estimation of its starting time sets the cooling regime somewhere in the transition between fast and
slow cooling. Giblin pointed out that the starting time of the tail could happen before 32 s, however the fact that the tail could start at $\sim$ 14 s implies that there could be a causality relationship between tail and HE PL component.

Surprisingly, also almost all the other bursts characterized by an extra HE PL component present a tail (see Table \ref{tab:tail}); the only exception is GRB 090902B, for which no tail has been reported in the literature, although a weak emission similar to a tail is evident after the T90 time interval. This suggests that the HE component and the tail could be related. 

\citet{fra12} investigated this possibility; in particular, they proposed for GRB980923 a unified model in which the tail can be understood as the early gamma-ray afterglow from forward shock Synchrotron emission, while the HE component arises from SSC  from the reverse shock. They found that this model accounts for the main characteristics of the burst: fluxes, break energies, starting times and spectral indices, provided that the ejecta is highly magnetized. An extension of this model has been recently applied with success to GRB090926A \citep{sacahui12}.

\section{Conclusion}\label{sec:conclusions} 

We have studied the prompt keV-MeV emission of GRB980923 as observed
by BATSE and TASC. We have shown that GRB980923 presents three different spectral
components. The main component described by a Band function with peak
energy of the order of 400 keV, a tail component described by a SBPL 
 with a break energy of the order of 200 keV and a HE 
component described by a PL. Table \ref{Resumen} summarizes the
main characteristics of each spectral component. The duration of this HE component is shorter than the duration of the burst itself, which is different compared with the other GRBs that presents a HE component distinct from the Band function. Also this HE component is observed in the middle of the burst episode, not at the end or beginning. A HE emission with a much shorter duration than the prompt emission clearly challenges the actual models, where a HE component has to last as long as the prompt phase or longer.


\begin{table}
\centering
\resizebox{0.95\textwidth}{!}{
\begin{tabular}{lcccccc}\hline \hline
{\bf Time from BATSE trigger (s)} &
{\bf 0 to 13} &
{\bf 13 to 33} &
{\bf 13 to 19.5} &
{\bf 19.5 to 21.5} &
{\bf 21.5 to 33}\\
\hline \hline
\multicolumn{6}{l}{\bf Band GRB function parameters} \\ \hline
A($\rm\frac{\times 10^{-2}ph}{s\; cm^{-2}keV}$) & $14.6^{+0.1}_{-0.2}$ & $24.7^{+0.2}_{-0.3}$ & $12.3^{+0.2}_{-0.2}$ & $29.5^{+0.1}_{-0.1}$   &  $32.1^{+0.3}_{-0.2}$ \\

$E_{peak}$(keV)  & $345.^{+4}_{-4}$ & $371^{+3}_{-3}$ &
$419^{+9}_{-9}$ & $419^{+15}_{-13}$ &  $346^{+3}_{-3}$\\

$\alpha$         & $-0.476^{+0.01}_{-0.01}$ & $-0.577^{+0.002}_{-0.002}$  & $-0.549^{+0.02}_{-0.021}$  & $-0.554^{+0.09}_{-0.091}$ &  $-0.563^{+0.014}_{-0.007}$\\

$\beta$          & $-2.64^{+0.04}_{-0.04}$ & $-3.06^{+0.09}_{-0.11}$ & $-2.48^{+0.07}_{-0.07}$ & $-4.01^{+0.97}_{-1.03}$ &  $-3.08^{+0.07}_{-0.07}$\\ 
 \hline
\multicolumn{6}{l}{\bf High-energy power law parameters}\\ \hline
A$_{PL}$($\rm\frac{\times 10^{-5}ph}{s\;cm^{-2}keV}$)& --- & $0.7^{+0.2}_{-0.2}$ & --- & $5.6^{+0.8}_{-0.8}$ & --- \\

$\gamma$       & ---   & $-1.28^{+0.26}_{-0.33}$ & ---  & $-1.44^{+0.07}_{-0.07}$ & --- \\
Probability
& ---  & $6.20\times10^{-7}$  & --- & $8.12\times10^{-11}$  & ---  \\ \hline
\multicolumn{6}{l}{\bf Normalization Factors}\\ \hline
SD7
& 1.28 & 1.31 & 1.21 & 1.44  & 1.36 \\
SD3     
& 1.23 & 1.30 & 1.05 & 1.51 & 1.39 \\
TASC 
& 0.76 & 0.94 & ---  &  --- & ---  \\
\hline \hline
\end{tabular}}
\caption[Spectral fitting parameters of the differential photon flux of GRB980923.]
{Spectral fitting parameters of the differential photon flux.
The first six rows contain the best spectral fit parameters, for different time intervals and including (columns 2 and 3) or not (columns 4, 5 and 6) TASC data. The seventh row shows the probability that the improvement in $\chi^{2}$ from the addition of the HE PL in the fit is due to chance, as determined by the $\chi^2$ test. The last three rows are the normalization factors relative to LAD7 (see secs. \ref{sec:analysis} and \ref{sec:results}) used in the fit.}
\label{tbl980923}
\end{table}

\begin{table}
\centering
\begin{tabular}{llccccc}\hline \hline
 & {\bf GRB name}       & {\bf HE duration (s)$^{(a)}$}& {\bf PL index$^{(b)}$} & {\bf Tail$^{(c)}$} \\ \hline \hline
(1) & GRB 980923    & 2 &-1.44 $\pm$ 0.07 &  yes\\
(2) & GRB 941017   & $\sim 200 $&$\sim$ -1 & yes  \\ 
(3) & GRB 090510   & $\sim$ 200 & -1.62 $\pm$ 0.03& yes \\ 
(4) & GRB 090902B & $\sim$ 1000 & -1.93 $\pm$ 0.01 & no\\ 
(5) & GRB 090926A & $\sim$ 4800 & -1.79 $\pm$ 0.02& yes\\ \hline \hline
\end{tabular}
\caption[GRB with extra HE PL component]
{Characteristics of the GRBs for which an extra HE PL component has been identified: duration of the HE emission, spectral index of the extra HE PL required to fit the prompt emission and presence of a tail. \\
$^{(a)}$ References for the HE duration: (1) see text; (2) \citet{GO03}; (3) \citet{Ackermann090510}; (4) \citet{Abdo090902B}; (5) \citet{Ackermann090926A}. \\$^{(b)}$ References for the  PL spectral index: (1) see text; (2) \citet{GO03}; (3) \citet{Ackermann090510}; (4) \citet{Abdo090902B};  (5) \citet{Ackermann090926A}.\\ $^{(c)}$ References for the tail: (1) \citet{GI99}; (2) \citet{GO05}; (3) \citet{GCN9336}; (4) - ; (5) \citet{GCN9951}, \citet{GCN9959}.}
\label{tab:tail}
\end{table}

\begin{table}
\centering
\resizebox{0.95\textwidth}{!}{
\begin{tabular}{lllllll}\hline \hline
Spectral component & Photon Flux Function & Peak Energy & Break Energy & Duration
&Start Time & Model \\ \hline \hline
main prompt emission & Band function & 400 keV & - & 33 s & 0 &Internal shocks \\
tail & SBPL &  & $\sim$ 200 keV & 400 s & 14 s & Synchrotron External Shocks \\
high energy component & PL & $>$200 MeV &  - & 2 s & 19.5 s & ? \\
\hline \hline
\end{tabular}}
\caption[Spectral components observed in GRB980923.]
{Summary of spectral components observed in GRB980923 and the main characteristics and theoretical models to explain them.}
\label{Resumen}
\end{table}

\acknowledgments

This work was supported by Consejo Nacional de Ciencia
y Tecnologia (grant number 103520) and UNAM (grant number PAPIIT IN105211).


\begin{thebibliography}{}
            
                                                               
\bibitem[Abdo et al.(2009)]{Abdo090902B} Abdo, A. A., et al. 2009,  \apj, 706, 138
                                
\bibitem[Ackermann et al.(2010)]{Ackermann090510}  Ackermann, M., et al. 2010, \apj, 716, 1178
 
\bibitem[Ackermann et al.(2011)]{Ackermann090926A} Ackermann, M., et al. 2011, \apj, 729, 114
 
\bibitem[Asano et al.(2009)]{asano09} Asano, K., Guiriec, S. \& Meszaros, P. 2009, \apj, 705, L191
 
\bibitem[Asano et al.(2011)]{asano11} Asano, K., et al. 2011, arXiv:1111.0127v1
 
\bibitem[Atwood et al.(2009)]{Atwood09} Atwood, W. B., et al. 2009, \apj, 697, 1071
 
\bibitem[Band et al.(1993)]{B93} Band, D., et al. 1993, \apj, 413, 281   
 
\bibitem[B\"ottcher \& Dermer(1998)]{bottcher98} B\"ottcher, M. \& Dermer, C. D. 1998, \apj, 499, L131 
 
\bibitem[Chiang \& Dermer(1999)]{chi99} Chiang, J. \& Dermer, C. D. 1999, \apj, 512, 699
 
\bibitem[de Pasquale et al.(2010)]{dePasquale2010} de Pasquale, F., et al. 2010, \apj, 709, L146
 
\bibitem[Dermer et al.(2000a)]{der00a} Dermer, C. D., B\"ottcher, M. \&  Chiang, J. 2000a, \apj, 537, 255
 
\bibitem[Dermer et al.(2000b)]{der00b} Dermer, C. D., Chiang, J. \&  Mitman, K. E.   2000b, \apj, 537, 785

\bibitem[Dermer \& Atoyan(2004)]{dermer04} Dermer, C. D. \& Atoyan, A. 2004, A\&A, 418, L5
 
\bibitem[Fishman et al.(1989)]{FI89} Fishman, G. J., et al. 1989, In: GRO Science Workshop Proceedings, GSFC, 3

\bibitem[Fraija et al.(2012)]{fra12} Fraija, N., Gonz\'alez, M. M. \& Lee, W. H.  2012, \apj,  751, 33
 
\bibitem[Giblin et al.(1999)]{GI99} Giblin, T. W., et al. 1999, \apj, 524, L47
                    
\bibitem[Golenetskii et al.(2009)]{GCN9959} Golenetskii, S., et al. 2009, GCN Circular 9959

\bibitem[Gonz\'alez et al.(2003)]{GO03} Gonz\'alez, M. M., et al. 2003, \nat, 424, 749

\bibitem[Gonz\'alez(2005)]{GO05} Gonz\'alez, M. M., Ph.D. Dissertation, U. of Wisconsin-Madison, 2005

\bibitem[Gonz\'alez et al.(2009)]{GO09} Gonz\'alez, M. M., et al. 2009, \apj, 696, 2155

\bibitem[Granot \& Guetta(2003)]{Granot03} Granot, J. \& Guetta, D.  2003, \apj, 598, L11
                                       
\bibitem[Granot(2010)]{Granot10} Granot, J. 2010, arXiv:1003.2452

\bibitem[Guiriec et al.(2010)]{GCN9336} Guiriec, S., Connaughton, V. \& Briggs, M. 2010, GCN Circular 9336

\bibitem[He et  al.(2011)]{he11}He, H.-N., et  al. 2011, \apj, 733, 22

\bibitem[Kaneko et al.(2006)]{K06} Kaneko, Y., et al. 2006, ApJSS, 166, 298
 
\bibitem[Kaneko et al.(2008)]{KA08} Kaneko, Y., et al. 2008, \apj, 677, 1168  

\bibitem[Kouveliotou et al.(1993)]{Kouv93} Kouveliotou, C., et al. 1993, \apj, 413, 101

\bibitem[Liu \& Wang(2011)]{liu11}Liu, R.   \& Wang,  X. 2011,  \apj, 730, 1

\bibitem[Meegan et al.(2009)]{Megaan09} Meegan, C., et al. 2009, \apj, 702, 791

\bibitem[Noda et al.(2009)]{GCN9951} Noda, K., et al. 2009, GRB Coordinates Network, 9951, 1

\bibitem[Panaitescu \& Kumar(2000)]{pan00b}Panaitescu, A. \& Kumar, P. 2000, \apj, 543, L66

\bibitem[Panaitescu \& M\'esz\'aros(1998)]{pan98}Panaitescu, A. \& M\'esz\'aros, P. 1998, \apj, 501, 772

\bibitem[Panaitescu \& M\'esz\'aros(2000)]{pan00a}Panaitescu, A. \& M\'esz\'aros, P. 2000, \apj, 544, 17    

\bibitem[Papathanassiou \& M\'esz\'aros(1996)]{pap96}Papathanassiou, H. \& M\'esz\'aros, P. 1996, \apj, 471, 91             

\bibitem[Pe'er \& Wijers(2006)]{pee06}  Pe'er,  A. \& Wijers, R. 2006, \apj, 643, 1036
       
\bibitem[Pendleton et al.(1995)]{PE95} Pendleton, G. N., et al. 1995, NIMPA, 364, 567

\bibitem[Pilla \& Loeb(1998)]{pil98}Pilla, R. \& Loeb, A. 1998, \apj, 494, 167

\bibitem[Preece(2000)]{PR00} Preece, R. D., et al. 2000, \apjs, 126, 19
                    
\bibitem[Razzaque, et al.(2009a)]{razzaque09} Razzaque, S., et al. 2009a, AIP Conf. Proc., 1133, 328

\bibitem[Razzaque et al.(2009b)]{razzaque09b} Razzaque, S., et al. 2009b, OAJ, 3, 150  

\bibitem[Sacahui et al.(2012)]{sacahui12} Sacahui, J. R., et al. 2012, arXiv:1203.1577              
                    
\bibitem[Sari  \& Esin(2001)]{sar01}Sari, R. \& Esin, A. A. 2001, \apj, 548, 787
                    
\bibitem[Sari  et  al.(1996)]{sar96}Sari, R., Narayan, R. \& Piran, T.  1996, \apj, 473, 204

\bibitem[Totani(1998a)]{tot98}Totani, T. 1998a, \apj, 502, L13  

\bibitem[Totani(1998b)]{tot98b} Totani, T. 1998b, \apj, 509, L81        

\bibitem[Veres \& M\'esz\'aros(2012)]{ver12} Veres, P.  \& M\'esz\'aros, P.  2012,  arXiv:1202.2821      

\bibitem[Vietri(1997)]{vietri97} Vietri, M. 1997, PRL, 78, 4328 

\bibitem[Wang et al.(2001a)]{wan01a}Wang, X. Y., Dai, Z. G. \& Lu, T. 2001a, \apj, 546, L33

\bibitem[Wang et al.(2001b)]{wan01b}Wang, X. Y., Dai, Z. G. \& Lu, T. 2001b, \apj, 556, 1010

\bibitem[Waxman(1997)]{wax97}Waxman, E. 1997, \apj, 489, L33

\bibitem[Waxman \& Bahcall(1997)]{wax97b} Waxman, E. \& Bahcall, J. 1997, PRL, 78, 2292 

\bibitem[Wei \& Lu(1998)]{wei98}Wei, D. M. \& Lu, T. 1998,  \apj, 505, 252

\end{thebibliography}
\end{document}